\documentclass[english,twocolumn,showpacs,preprintnumbers,prl]{revtex4}
\usepackage[T1]{fontenc}
\usepackage[latin1]{inputenc}
\usepackage{amssymb}

\makeatletter

\providecommand{\boldsymbol}[1]{\mbox{\boldmath $#1$}}

\usepackage{graphics}

\usepackage{psfig}
\usepackage{color}

\usepackage{babel}
\makeatother
\begin{document}

\title{Simpler Variational Problem for Statistical Equilibria of the 2D
Euler Equation and Other Systems with Long Range Interactions}

\author{Freddy Bouchet}

\email{Freddy.Bouchet@inln.cnrs.fr}

\affiliation{Institut Non Linéaire de Nice, INLN, CNRS, UNSA, 1361 route des lucioles,
06 560 Valbonne - Sophia Antipolis, France}

\date{{\normalsize \today}}

\begin{abstract}
The Robert-Sommeria-Miller equilibrium statistical mechanics predicts
the final organization of two dimensional flows. This powerful theory
is difficult to handle practically, due to the complexity associated
with an infinite number of constraints. Several alternative simpler
variational problems, based on Casimir's or stream function functionals,
have been considered recently. We establish the relations between
all these variational problems, justifying the use of simpler formulations.
\end{abstract}


\maketitle
We consider the 2D Euler equation, on a domain $\mathcal{D}$ \begin{equation}
\frac{\partial\omega}{\partial t}+\mathbf{v}\boldsymbol{\cdot\nabla}\omega=0\,\,\mathbf{;\,\, v}=\mathbf{e}_{z}\times\boldsymbol{\nabla}\psi\,\,;\,\,\omega=\Delta\psi\label{Eq:advection}\end{equation}
 where $\omega$ is the vorticity, $\mathbf{v}$ the velocity and
$\psi$ the stream function ($\mbox{with}\,\,\,\psi=0\,\,\,\mbox{on}\,\,\,\partial\mathcal{D}$,
$\mathcal{D}$ is simply connected).

The equilibrium statistical mechanics of the 2D Euler equation (the
Robert-Sommeria-Miller (RSM) theory \cite{Robert:1991_JSP_Meca_Stat,Robert_Sommeria_1991JFM,Miller:1990_PRL_Meca_Stat}),
assuming ergodicity, predicts the final organization of the flow,
on a coarse grained level (see \cite{Eyink_Sreenivasan_2006_Rev_Modern_Physics}
for a recent review of Onsager ideas, that inspired the RSM theory).
Besides its elegance, this predictive theory is a very interesting
and useful scientific tool. 

From a mathematical point of view, one has to solve a microcanonical
variational problem (MVP) : maximizing a mixing entropy $\mathcal{S}[\rho]=-\int_{\mathcal{D}}d^{2}x\int d\sigma\,\rho\log\rho,$
with constraints on energy $E$ and vorticity distribution $d$ \[
S(E_{0},d)=\hspace{-0.3cm}\sup_{\left\{ \rho|N\left[\rho\right]=1\right\} }\hspace{-0.3cm}\left\{ \mathcal{S}[\rho]\ |\ E\left[\overline{\omega}\right]=E_{0}\ ,D\left[\rho\right]=d\ \right\} \,\,\mbox{(MVP).}\]
 $\rho\left(\mathbf{x},\sigma\right)$ depends on space $\mathbf{x}$
and vorticity $\sigma$ variables. 

The theoretical predictability of RSM theory requires the knowledge
of all conserved quantities. The infinite number of Casimir's functionals
(this is equivalent to vorticity distribution $d$) have then to be
considered. This is a huge practical limitation. When faced with real
flows, physicists can then either give physical arguments for a given
type of distribution $d$ (modeler approach) or ask whether it exists
some distribution $d$ with RSM equilibria close to the observed flow
(inverse problem approach). However, in any cases the complexity remains
: the class of RSM equilibria is huge.

During recent years, authors have proposed alternative approaches,
which led to practical and/or mathematical simplifications in the
study of such equilibria. As a first example, Ellis, Haven and Turkington
\cite{EllisHavenTurkington:2002_Nonlinearity_Stability} proposed
to treat the vorticity distribution canonically (in a canonical statistical
ensemble). From a physical point of view, a canonical ensemble for
the vorticity distribution would mean that the system is in equilibrium
with a bath providing a prior distribution of vorticity. As such a
bath does not exist, the physically relevant ensemble remains the
one based on the dynamics : the microcanonical one. However, the Ellis-Haven-Turkington
approach is extremely interesting as it provides a drastic mathematical
and practical simplification to the problem of computing equilibrium
states. A second example, largely popularized by Chavanis \cite{Chavanis_Generalized_Entropy_2003,Chavanis_2005PhyD_Generalizedentropy},
is the maximization of generalized entropies. Both the prior distribution
approach of Ellis, Haven and Turkington or its generalized thermodynamics
interpretation by Chavanis lead to a second variational problem: the
maximization of Casimir's functionals, with energy constraint (CVP)
\[
C(E_{0},s)=\inf_{\omega}\left\{ \mathcal{C}_{s}[\omega]=\int_{\mathcal{D}}s(\omega)d^{2}x\ |\ E\left[\omega\right]=E_{0}\ \right\} \,\,\mbox{(CVP)}\]
where $\mathcal{C}_{s}$ are Casimir's functionals, and $s$ a convex
function (Energy-Casimir functionals are used in classical works on
nonlinear stability of Euler stationary flows \cite{Arnold_1966,Holm_Marsden_Ratiu_Weinstein_1985PhysRev},
and have been used to show the nonlinear stability of some of RSM
equilibrium states \cite{Robert_Sommeria_1991JFM,Michel_Robert_1994JFM}). 

Another class of variational problems (SFVP), that involve the stream
function only (and not the vorticity), has been considered in relation
with the RSM theory\[
D\left(G\right)=\inf_{\psi}\left\{ \int_{\mathcal{D}}d^{2}x\,\left[-\frac{1}{2}\left|\nabla\psi\right|^{2}+G\left(\psi\right)\right]\ \right\} \,\,\mbox{(SFVP)}\]
 Such (SFVP) functionals have been used to prove the existence of
solutions to the equation describing critical points of (MVP) \cite{Michel_Robert_1994JFM}.
Interestingly, for the Quasi-geostrophic model, in the limit of small
Rossby deformation radius, such a SFVP functional is similar to the
Van-Der-Walls Cahn Hilliard model which describes phase coexistence
in usual thermodynamics \cite{Bouchet_Sommeria_2002_JFM_GRS,Bouchet_These}.
This physical analogy has been used to make precise predictions in
order to model Jovian vortices \cite{Bouchet_Sommeria_2002_JFM_GRS,Bouchet_Dumont_Jupiter}.
(SFVP) functionals are much more regular than (CVP) functionals and
thus also very interesting for mathematical purposes. 

When we prescribe appropriate relations between the distribution function
$d$, the functions $s$ and $G$, the three previous variational
problems have the same critical points. This has been one of the motivations
for their use in previous works. However, a clear description of the
relations between the stability of these critical points is still
missing (is a (CVP) minimizer a RSM equilibria, or does a RSM equilibria
minimize (CVP) ?). This has led to fuzzy discussions in recent papers.
Providing an answer is a very important theoretical issue because,
as explained previously, it will lead to deep mathematical simplifications
and will provide useful physical analogies.

The aim of this short paper is to establish the relation between these
three variational problems. The result is that any minimizer (global
or local) of (SFVP) minimizes (CVP) and that any minimizer of (CVP)
is a RSM equilibria. The opposite statements are wrong in general.
For instance (CVP) minimizers may not minimize (SFVP), but be only
saddles. Similarly, RSM equilibria may not minimize (CVP) but be only
saddles, even if no explicit example has yet been exhibited. 

These results have several interesting consequences :

\begin{enumerate}
\item As the ensemble of (CVP) minimizers is a sub-ensemble of the ensemble
of RSM equilibria, one can not claim that (CVP) are more relevant
for applications than RSM equilibria.
\item The link between (CVP) and RSM equilibria provides a further justification
for studying (CVP). 
\item Based on statistical mechanics arguments, when looking at the Euler
evolution on a coarse-grained level, it may be natural to expect the
RSM entropy to increase. There is however no reason to expect such
a property to be true for the Casimir's functional. As explained above,
it may also happen that entropy extrema being (CVP) saddles. 
\end{enumerate}
In order to simplify the discussion, we keep only the energy constraint
at the level of the Casimir functional (CVP). Adding other constraints,
such as the circulation \cite{EllisHavenTurkington:2000_Inequivalence},
or even the microscopic enstrophy, does not change the discussion.

We note that all the discussion can be easily generalized to any system
with long range interactions (self-gravitating systems, Vlasov Poisson
system) \cite{DRAW:2002_Houches}.\\

In the first section, we explain the link between a constrained variational
problem and its relaxed version. We explain that any minimizer of
the second is a minimizer of the first. In the second section, we
present the microcanonical variational problem (MVP). We then introduce
a mixed grand canonical ensemble by relaxing the vorticity distribution
constraint in the RSM formalism. We prove in the third section that
this mixed ensemble is equivalent to (CVP). Similarly, in the last
section we prove that (SFVP) variational problem is equivalent to
a relaxed version of (CVP).

\section{Relations between constrained and relaxed variational problems\label{sec:contrainte}}

We discuss briefly relations between a constrained variational problem
and its relaxed version. This situation appears very often in statistical
mechanics when passing from one statistical ensemble to another. We
assume that the Lagrange's multipliers rule applies. Let us consider
the two variational problems {\footnotesize \[
G(C)=\inf_{x}\left\{ g(x)|c(x)=C\right\} \,\,\,\mbox{and\,\,\,}H(\gamma)=\inf_{x}\left\{ h_{\gamma}(x)=g(x)-\gamma c(x)\right\} .\]
}$G$ is the constrained variational problem and $H$ is the relaxed
one, $\gamma$ is the Lagrange multiplier (or the dual variable) associated
to $C$. We have the results :

\begin{enumerate}
\item {\footnotesize $H\left(\gamma\right)=\inf_{C}\left\{ G(C)-\gamma C\right\} $}
and {\footnotesize $G(C)\geq\sup_{\gamma}\left\{ \gamma C+H(\gamma)\right\} $}{\footnotesize \par}
\item If $x_{m}$ is a minimizer of $h_{\gamma}$ then $x_{m}$ is also
a minimizer of $G(C)$ with the constraint $C=c(x_{m})$
\item If $x_{m}$ is a minimizer of $G(C)$, then it exists a value of $\gamma$
such that $x_{m}$ is a critical point of $h_{\gamma}$, but $x_{m}$
may not be a minimizer of $h_{\gamma}$ but just a saddle. Then $x_{m}$
is a minimizer of $h_{\gamma}$ if and only if {\footnotesize $G(C)=\sup_{\gamma}\left\{ H(\gamma)+\gamma C\right\} $}
if and only if $G(C)$ coincides with the convex hull of $G$ in $C$.
In this last situation the two variational problems are said \emph{equivalent}. 
\end{enumerate}
Such results are extremely classical. More detailed results in this
context may be found in \cite{EllisHavenTurkington:2000_Inequivalence}.
Situations of ensemble inequivalence have been classified, in relation
with phase transitions \cite{Bouchet_Barre_2005_JSP_Classification}. 

Equality in point \textbf{1.} follows from the remark that {\footnotesize \[
\hspace{-0.3cm}H(\gamma)=\inf_{C}\left\{ \inf_{x}\left\{ g(x)-\gamma c(x)|c(x)=C\right\} \right\} =\inf_{C}\left\{ \inf_{x}\left\{ g(x)|c(x)=C\right\} -\gamma C\right\} .\]
}We remark that $-H$ is the Legendre-Fenchel transform of $G$. The
inequality of point \textbf{1.} is then a classical convex analysis
result. We have for any value of $\gamma$, {\footnotesize \begin{equation}
\begin{array}{cc}
G(C) & =\inf_{x}\left\{ g(x)|c(x)=C\right\} =\inf_{x}\left\{ g(x)-\gamma c(x)|c(x)=C\right\} +\gamma C\\
 & \geq\inf_{x}\left\{ g(x)-\gamma c(x)\right\} +\gamma C=H(\gamma)+\gamma C.\end{array}\label{eq:inequlity}\end{equation}
}This is a direct proof of the inequality of point 1.

Point \textbf{2.} : for $x_{m}$ a minimizer of $h_{\gamma}$ and
$x$ with $c(x)=c(x_{m})$, we have $g(x_{m})=h_{\gamma}(x_{m})+\gamma c(x_{m})\leq h_{\gamma}\left(x\right)+\gamma c(x_{m})=g(x)$.
This proves \textbf{2.}. First assertion of \textbf{3.} is Lagrange's
multipliers rule. Clearly, $x_{m}$ is a minimizer of $h_{\gamma}$
if and only if equality occurs in (\ref{eq:inequlity}). It is a classical
result of convex analysis that the convex hull of $G$ is the Legendre-Fenchel
transform of $-H$. This concludes the proof of \textbf{3.}. Many
examples where $x_{m}$ is a saddle may be found in the literature
(see \cite{Bouchet_Barre_2005_JSP_Classification}, or examples in
the context of Euler equation in \cite{Smith_ONeil_Physics_Fluids_1990_Inequivalence,Kiessling_1993_PointVortex_CommPureApplMaths,CagliotiLMP:1995_CMP_II(Inequivalence)}).

\section{RSM statistical mechanics  \label{sec:Constrained-entropy-maxima}}

Euler's equation (\ref{Eq:advection}) conserves the kinetic energy
\begin{equation}
E\left[\omega\right]=\frac{1}{2}\int_{\mathcal{D}}d^{2}x\,\left(\boldsymbol{\nabla}\psi\right)^{2}=-\frac{1}{2}\int_{\mathcal{D}}d^{2}x\,\omega\psi=E_{0}\label{Eq:energie}\end{equation}
 and for sufficiently regular functions $s$, Casimirs\begin{equation}
\mathcal{C}_{s}[\omega]=\int_{\mathcal{D}}d^{2}x\, s(\omega).\label{eq:casimir}\end{equation}
Let us define $A\left(\sigma\right)$ the area of $\mathcal{D}$ with
vorticity values lower than $\sigma$, and $d\left(\sigma\right)$
the vorticity distribution\begin{equation}
d\left(\sigma\right)=\frac{1}{\left|\mathcal{D}\right|}\frac{dA}{d\sigma}\,\,\,\mbox{with\,\,\,}A\left(\sigma\right)=\int_{\mathcal{D}}d^{2}x\,\chi_{\left\{ \omega\left({\bf x}\right)\leq\sigma\right\} },\label{eq:distribution_vorticite}\end{equation}
where $\chi_{B}$ is the characteristic function of ensemble $B$,
and $\left|\mathcal{D}\right|$ is the area of $\mathcal{D}$. As
Euler's equation (\ref{Eq:advection}) is a transport equation by
an incompressible flow, $d\left(\sigma\right)$ (or equivalently $A\left(\sigma\right)$)
is conserved by the dynamics. Conservation of distribution $d\left(\sigma\right)$
and of all Casimir's functionals (\ref{eq:casimir}) is equivalent.

\subsection{RSM microcanonical equilibria (MVP)}

We present the classical derivation \cite{Robert_Sommeria_1991JFM}
of the microcanonical variational problem which describes RSM equilibria.
Such equilibria describe the most probable mixing of the vorticity
$\omega$, constrained by the vorticity distribution (\ref{eq:distribution_vorticite})
and energy (\ref{Eq:energie}) (other conservation laws could be considered,
for instance if the domain $D$ has symmetries). 

We make a probabilistic description of the flow. We define $\rho\left(\sigma,\mathbf{x}\right)$
the local probability that the microscopic vorticity $\omega$ take
a value $\omega\left({\bf \mathbf{x}}\right)=\sigma$ at position
$\mathbf{x}$. As $\rho$ is a local probability, it verifies a local
normalization\begin{equation}
N\left[\rho\right](\mathbf{x})\equiv\int_{-\infty}^{+\infty}\hspace{-0.5cm}d\sigma\,\,\rho\left(\sigma,\mathbf{x}\right)=1.\label{eq:normalisation}\end{equation}
The known vorticity distribution (\ref{eq:distribution_vorticite})
imposes \begin{equation}
D\left[\rho\right](\sigma)\equiv\int_{\mathcal{D}}d\mathbf{x}\,\rho\left(\sigma,{\bf \mathbf{x}}\right)=d\left(\sigma\right).\label{eq:distribution_ro}\end{equation}
We are interested on a locally averaged, coarse-grained description
of the flow. The coarse grained vorticity is \begin{equation}
\overline{\omega}\left(\mathbf{x}\right)=\int_{-\infty}^{+\infty}\hspace{-0.5cm}d\sigma\,\,\sigma\rho\left(\sigma,\mathbf{x}\right).\label{eq:vorticite_coarse_grained}\end{equation}
$\overline{\psi}=\Delta\bar{\omega}$ is the coarse grained stream
function. The energy may be expressed in terms of the coarse-grained
vorticity distribution as\begin{equation}
E\left[\overline{\omega}\right]\equiv-\frac{1}{2}\int_{\mathcal{D}}\overline{\psi}\overline{\omega}d\mathbf{x}\simeq E_{0}.\label{eq:energie_coarse_grained}\end{equation}

The entropy is a measure of the number of microscopic vorticity fields
which are compatible with a distribution $\rho$. By classical arguments,
such a measure is given by the Maxwell-Boltzmann entropy\begin{equation}
\mathcal{S}\left[\rho\right]\equiv-\int_{\mathcal{D}}d^{2}x\int_{-\infty}^{+\infty}\hspace{-0.5cm}d\sigma\,\rho\log\rho.\label{eq:entropie_Maxwell_Boltzmann}\end{equation}
The most probable mixing for the potential vorticity is thus given
by the probability $\rho_{eq}$ which maximizes the Maxwell-Boltzmann
entropy (\ref{eq:entropie_Maxwell_Boltzmann}), subject to the three
constraints (\ref{eq:normalisation},\ref{eq:distribution_ro} and
\ref{eq:energie_coarse_grained}). The equilibrium entropy $S(E_{0},d)$,
the value of the constrained entropy maxima, is then given by the
microcanonical variational problem (MVP) (see the introduction).

Using the Lagrange's multipliers rule, there exists $\beta$ and $\alpha\left(\sigma\right)$
(the Lagrange parameters associated to the energy and vorticity distribution,
respectively) such that the critical points of (MVP) verify\begin{equation}
\rho_{eq}\left(\mathbf{x},\sigma\right)=\frac{1}{z_{\alpha}\left(\beta\psi_{eq}\right)}\exp\left[\sigma\beta\psi_{eq}-\alpha\left(\sigma\right)\right],\label{eq:ro_equilibre}\end{equation}
where we define \begin{equation}
z_{\alpha}\left(u\right)=\int_{-\infty}^{+\infty}\hspace{-0.5cm}d\sigma\,\exp\left[\sigma u-\alpha\left(\sigma\right)\right]\,\,\,\mbox{and}\,\,\, f_{\alpha}\left(u\right)=\frac{d}{du}\log z_{\alpha}.\label{eq:definition_z_alpha}\end{equation}
We note that $z_{\alpha}$ is positive, $\log z_{\alpha}$ is convex,
and thus $f_{\alpha}$ is strictly increasing.

From (\ref{eq:ro_equilibre}), using (\ref{eq:vorticite_coarse_grained}),
the equilibrium vorticity is \begin{equation}
\omega_{eq}=f_{\alpha}\left(\beta\psi_{eq}\right)\,\,\,\mbox{or equivalently}\,\,\, g_{\alpha}\left(\omega_{eq}\right)=\beta\psi_{eq},\label{eq:vorticite_equilibre}\end{equation}
where $g_{\alpha}$ is the inverse of $f_{\alpha}$. The actual equilibrium
$\omega_{eq}$ is the minimizer of the entropy while verifying the
constraints, between all critical points for any possible values of
$\beta$ and $\alpha$.

We note that solutions to (\ref{eq:vorticite_equilibre}) are stationary
flows.

\subsection{RSM constrained grand canonical ensemble }

We consider the statistical equilibrium variational problem (MVP),
but we relax the vorticity distribution constraint. This constrained
(or mixed) grand canonical variational problem is \begin{equation}
G(E_{0},\alpha)=\inf_{\hspace{-0.1cm}\inf_{\left\{ \rho|N\left[\rho\right]=1\right\} }\hspace{-0.1cm}}\left\{ \mathcal{G}_{\alpha}[\rho]\ \big|\ E\left[\overline{\omega}\right]=E_{0}\right\} ,\label{eq:probleme_variationnel_grand_canonique}\end{equation}
with the Gibbs potential functional defined as \[
\mathcal{G}_{\alpha}\left[\rho\right]\equiv-S\left[\rho\right]+\int_{\mathcal{D}}d^{2}x\int_{-\infty}^{+\infty}\hspace{-0.5cm}d\sigma\,\alpha\left(\sigma\right)\rho\left(\mathbf{x},\sigma\right).\]

In the following section, we prove that (\ref{eq:probleme_variationnel_grand_canonique})
is equivalent to the constraint Casimir one (CVP). Using the results
of the first section, relating constrained and relaxed variational
problems, we can thus conclude that minimizers of (CVP) are RSM equilibria,
but the converse is wrong in general, as stated in the introduction.

\section{Constrained Casimir (CVP) and grand canonical ensembles are equivalent\label{sec:Variational_Problem_Casimirs}}

\subsection{Equivalence }

We consider a Casimir's functionnal (\ref{eq:casimir}), where $s$
is assumed to be convex. The critical points of the constrained Casimir
variational problem (CVP, see introduction) verify \begin{equation}
\frac{ds}{d\omega}\left(\omega_{eq}\right)=\beta\psi_{eq},\label{eq:equilibre_Casimirs}\end{equation}
where $\beta$ is the opposite of the Lagrange's multiplier for the
energy. Solutions to this equation are stationary states for the Euler
equation. Moreover, with suitable assumptions for the function $s$,
such flows are proved to be nonlinearly stable \cite{Arnold_1966}. 

This last equation is very similar to the one verified by RSM equilibria
(\ref{eq:vorticite_equilibre}). Indeed let us define $s_{\alpha}$
the Legendre-Fenchel transform of $\log z_{\alpha}$ \begin{equation}
s_{\alpha}\left(\omega\right)=\sup_{u}\left\{ u\omega-\log z_{\alpha}\left(u\right)\right\} .\label{eq:Legendre-Fenchel-s}\end{equation}
Then $s_{\alpha}$ is convex. Moreover, if $\log z_{\alpha}$ is differentiable,
then direct computations lead to \begin{equation}
s_{\alpha}\left(\omega\right)=\omega g_{\alpha}\left(\omega\right)-\log\left(z_{\alpha}\left(g_{\alpha}\left(\omega\right)\right)\right)\label{eq:s_alpha}\end{equation}
 and to $ds/d\omega=g_{\alpha}$. The equilibrium relations (\ref{eq:vorticite_equilibre}),
and (\ref{eq:equilibre_Casimirs}) with $s=s_{\alpha}$, are the same
ones. It been observed in the past by a number of authors \cite{Robert_Sommeria_1991JFM}.\\

Let us that (\ref{eq:probleme_variationnel_grand_canonique}) and
(CVP) are equivalent if $s=s_{\alpha}$. More precisely, we assume
that Lagrange's multipliers rule applies, and we prove that minimizers
of both variational problems have the same $\omega_{eq}$ and that
$C(E_{0},s_{\alpha})=G(E_{0},\alpha)$. 

We consider a minimizer $\rho_{eq}$ of (\ref{eq:probleme_variationnel_grand_canonique})
and $\omega_{eq}=\int d\sigma\,\sigma\rho_{eq}$. Then $E\left[\omega_{eq}\right]=E_{0}$
and $G(E_{0},\alpha)=\mathcal{G}_{\alpha}\left[\rho_{eq}\right]$.
A Lagrange multiplier $\beta$ then exists such that $\rho_{eq}$
verifies equation (\ref{eq:ro_equilibre}). Direct computation gives
$\rho_{eq}\log\rho_{eq}+\alpha\rho_{eq}=\exp\left(\beta\sigma\psi_{eq}-\alpha\left(\sigma\right)\right)\left[-\log z_{\alpha}\left(\beta\psi_{eq}\right)+\beta\sigma\psi_{eq}\right]/z_{\alpha}\left(\beta\psi_{eq}\right)$.
Using $\omega_{eq}=\int d\sigma\,\sigma\rho_{eq}$, (\ref{eq:vorticite_equilibre})
and (\ref{eq:s_alpha}), we obtain\begin{equation}
\hspace{-0.5cm}\int_{-\infty}^{+\infty}\hspace{-0.5cm}d\sigma\,\left(\rho_{eq}\log\rho_{eq}+\alpha\rho_{eq}\right)=-\log z_{\alpha}\left(\beta\psi_{eq}\right)+\beta\psi_{eq}\omega_{eq}=s_{\alpha}\left(\omega_{eq}\right).\label{eq:calcul_2}\end{equation}
From the definitions of $\mathcal{G}$ and $\mathcal{C}$, we obtain
$G(E_{0},\alpha)=\mathcal{G}_{\alpha}\left[\rho_{eq}\right]=\mathcal{C}_{s_{\alpha}}\left[\omega_{eq}\right]$.
Now, as $C$ is an infimum, $C_{s_{\alpha}}\left[\omega_{eq}\right]\geq C(E_{0},s_{\alpha})$
and\[
G(E_{0},\alpha)\geq C(E_{0},s_{\alpha}).\]
We now prove the opposite inequality. Let $\omega_{eq,2}$ be a minimizer
of (CVP) with $s=s_{\alpha}$. Then it exists $\beta_{2}$ such that
(\ref{eq:equilibre_Casimirs}) is verified with $ds_{\alpha}/d\omega=g_{\alpha}$.
We then define $\rho_{eq,2}\equiv\exp$$\left[\sigma\beta_{2}\psi_{eq,2}-\alpha\left(\sigma\right)\right]/z_{\alpha}\left(\beta_{2}\psi_{eq,2}\right)$.
Following the same computations as in (\ref{eq:calcul_2}) , we conclude
that $\mathcal{G}_{\alpha}\left[\rho_{eq,2}\right]=\mathcal{C}_{s_{\alpha}}\left[\omega_{eq,2}\right]=C(E_{0},s_{\alpha})$.
Then using that $G$ is an infimum we have $G(E_{0},\alpha)\leq C(E_{0},s_{\alpha})$
and thus\[
G(E_{0},\alpha)=C(E_{0},s_{\alpha}).\]
Then $C_{s_{\alpha}}\left[\omega_{eq}\right]=C(E_{0},s_{\alpha})=G(E_{0},\alpha)=\mathcal{G}_{\alpha}\left[\rho_{eq,2}\right]$.
Thus $\omega_{eq}$ and $\rho_{eq,2}$ are minimizer of (CVP) and
of (\ref{eq:probleme_variationnel_grand_canonique}) respectively.
But as such minimizers are in general not unique, $\omega_{eq}$ may
be different from $\omega_{eq,2}$ and $\beta$ may be different from
$\beta_{2}$.

A formal, but very instructive, alternative way to obtain equivalence
between (CVP) and (\ref{eq:probleme_variationnel_grand_canonique})
is to note that \begin{equation}
C_{s_{\alpha}}\left[\omega\right]=\hspace{-0.1cm}\inf_{\left\{ \rho|N\left[\rho\right]=1\right\} }\hspace{-0.1cm}\left\{ \mathcal{G}_{\alpha}\left[\rho\right]\big|\int_{-\infty}^{+\infty}\hspace{-0.5cm}d\sigma\,\sigma\rho=\omega(\mathbf{x})\right\} .\label{eq:intermediaire}\end{equation}
We do not detail the computation. A proof of this result is easy as
we minimize a convex functional with linear constraints. Then, from
(\ref{eq:probleme_variationnel_grand_canonique}), using (\ref{eq:intermediaire}),
we obtain {\footnotesize \[
\hspace{-0.4cm}G(E_{0},\alpha)=\inf_{\omega}\left\{ \hspace{-0.15cm}\inf_{\left\{ \rho|N\left[\rho\right]=1\right\} }\hspace{-0.15cm}\left\{ \mathcal{G}_{\alpha}[\rho]\ \big|\ \ \hspace{-0.2cm}\int_{-\infty}^{+\infty}\hspace{-0.5cm}d\sigma\,\sigma\rho=\omega(\mathbf{x})\right\} \big|\ E\left[\overline{\omega}\right]=E_{0}\right\} =C(E_{0},s_{\alpha}).\]
}{\footnotesize \par}

\subsection{Second variations and local stability equivalence}

In the previous section, we have proved that the constrained Casimir
(CVP) and mixed ensemble (\ref{eq:probleme_variationnel_grand_canonique})
variational problems are equivalent, for global minimization. Does
this equivalence also hold for local minima ? We now prove that the
reply is positive. 

We say that a critical point $\rho_{eq}$ of the constrained mixed
ensemble variational problem (\ref{eq:probleme_variationnel_grand_canonique})
is locally stable iff the second variations $\delta^{2}\mathcal{J}_{\alpha}$,
of the associated free energy $\mathcal{J}_{\alpha}=\mathcal{G}_{\alpha}+\beta E$,
are positive for perturbations $\delta\rho$ that respect the linearized
energy constraints $\int_{\mathcal{D}}\psi_{eq}\delta\omega=0$, where
$\delta\omega=\int d\sigma\,\sigma\delta\rho$. Similarly, the second
variations $\delta^{2}\mathcal{D}_{s}$ of the free energy $\mathcal{D}_{s}=\mathcal{C}_{s}+\beta E$
define the local stability of the Casimir maximization. 

By a direct computation, we have $\delta^{2}\mathcal{G}_{\alpha}\left[\delta\rho\right]=-\delta^{2}\mathcal{S}_{\alpha}\left[\delta\rho\right]=\int_{\mathcal{D}}d\mathbf{x}\int d\sigma\frac{1}{\rho_{eq}}\left(\delta\rho\right)^{2}$
and ${\delta^{2}\mathcal{C}}_{s_{\alpha}}\left[\delta\omega\right]=\int_{\mathcal{D}}d\mathbf{x}\, s_{\alpha}^{''}\left(\omega_{eq}\right)\left(\delta\omega\right)^{2}$.
We decompose any $\delta\rho$ as {\footnotesize \[
\delta\rho=\delta\rho^{\parallel}+\delta\rho^{\perp}\;\mbox{with}\;\delta\rho^{\parallel}=\frac{\delta\omega}{f_{\alpha}^{'}}\left(\frac{-z_{\alpha}^{'}+\sigma z_{\alpha}}{z_{\alpha}^{2}}\right)\exp\left[\sigma\beta\psi_{eq}-\alpha\left(\sigma\right)\right].\]
} In this expression, the functions $f_{\alpha}^{'}$, $z_{\alpha}$
and $z_{\alpha}^{'}$ are evaluated at the point $\beta\psi_{eq}$.
Using the definition of $f_{\alpha}$ and of $z_{\alpha}$ (\ref{eq:definition_z_alpha}),
and the fact that $f_{\alpha}^{'}=\left(-z_{\alpha}^{'2}+z_{\alpha}z_{\alpha}^{''}\right)/z_{\alpha}^{2}$
we easily verify that the above expression is consistent with the
relation $\delta\omega=\int d\sigma\,\sigma\delta\rho$. 

Moreover by lengthy but straightforward computations, we verify that
$\int d\sigma\,\delta\rho^{\parallel}\delta\rho^{\perp}/\rho_{eq}=0$.
In this sense, the decomposition $\delta\rho=\delta\rho^{\parallel}+\delta\rho^{\perp}$
distinguishes the variations of $\rho$ that are normal to equilibrium
relation (\ref{eq:ro_equilibre}) from the tangential ones.

From $s_{\alpha}^{'}=g_{\alpha}$ and using that $\left(g_{\alpha}\right)^{-1}=f_{\alpha}$,
we obtain $s_{\alpha}^{''}=\left(f_{\alpha}^{'}\right)^{-1}$. using
this relation we obtain $\int d\sigma\,\left(\delta\rho^{\parallel}\right)^{2}/\rho_{eq}=s_{\alpha}^{''}\left(\omega_{eq}\right)\left(\delta\omega\right)^{2}$.
We thus conclude \begin{equation}
\delta^{2}\mathcal{J}_{\alpha}\left[\delta\rho\right]=\int_{\mathcal{D}}d^{2}x\int_{-\infty}^{+\infty}\hspace{-0.5cm}d\sigma\,\,\frac{1}{\rho_{eq}}\left(\delta\rho^{\perp}\right)^{2}+{\delta^{2}\mathcal{D}}_{s_{\alpha}}\left[\delta\omega\right].\label{eq:variations_secondes}\end{equation}
To the best of our knowledge, this equality has never been derived
before. It may be very useful as second variations are involved in
many stability discussions. 

From equality (\ref{eq:variations_secondes}), it is obvious that
the second variations of $\mathcal{J}_{\alpha}$ are positive iff
the second variations of $\mathcal{D}_{s_{\alpha}}$ are positive.
If we also note that perturbations which respect the linearized energy
constraint are the same for both functionals, we conclude that the
local stabilities of the two variational problems are equivalent.

\section{Relation between  RSM equilibria and stream function functionals\label{sec:SFVP}}

In this section, we establish the relation between stream function
functionals and RSM equilibria. For this we consider the constrained
Casimir variational problem (CVP). However, we relax the energy constraint.
We thus consider the free energy associated to CVP \begin{equation}
F(\beta,s)=\inf_{\omega}\left\{ \mathcal{F}_{s}[\omega]=\mathcal{C}_{s}[\omega]\ +\beta E\left[\omega\right]\ \right\} .\label{eq:Energie_Libre}\end{equation}
This is an Energy-Casimir functional \cite{Arnold_1966}. As previously
explained, minima of this relaxed variational problem are also minimum
(CVP). It is thus also a RSM equilibria.

Let $G$ be the Legendre-Fenchel transform of the function $s$ :
$G(z)=\sup_{y}\left\{ zy-s(y)\right\} $. $G$ is thus convex. In
the following, we will show that the variational problem (\ref{eq:Energie_Libre})
is equivalent to the SFVP \[
D\left(\beta,G\right)=\inf_{\psi}\left\{ \mathcal{D}_{G}[\psi]=\int_{\mathcal{D}}d^{2}x\,\left[-\frac{\beta}{2}\left|\nabla\psi\right|^{2}-G\left(\beta\psi\right)\right]\ \right\} .\]
 More precisely we prove that 

\begin{enumerate}
\item $F\left(\beta,s\right)=D(\beta,G)$
\item $\omega_{eq}=\Delta\psi_{eq}$ is a global minimizer of $\mathcal{F}_{s}$
if and only if $\psi_{eq}$ is a global minimizer of $\mathcal{D}_{G}$ 
\item If $\psi_{eq}$ is a local minimizer of $\mathcal{F}_{s}$ then it
is a local minimizer of $\mathcal{D}_{s}$.
\end{enumerate}
In order to prove these results, it is sufficient to prove 

\begin{description}
\item [{a)}] $\omega_{c}=\Delta\psi_{c}$ is a critical points of $\mathcal{F}_{s}$
if and only if $\psi_{c}$ is a critical point of $\mathcal{D}_{G}$,
and then $\mathcal{F}_{s}\left[\omega_{c}\right]=\mathcal{D}_{G}\left[\psi_{c}\right]$
\item [{b)}] For any $\omega=\Delta\psi$ , $\mathcal{F}_{s}\left[\omega\right]\geq\mathcal{D}_{G}\left[\psi\right]$.
\end{description}
Point \textbf{a)} has been noticed in \cite{Bouchet_These}, and is
actually sufficient to prove points \textbf{1)} and \textbf{2)}. The
inequality \textbf{b)} \cite{Wolansky_Ghil_1998_CMaPh} proves that
$\mathcal{\mathcal{D}_{G}}$ is a support functional to $\mathcal{F}_{s}$
\cite{Wolansky_Ghil_1998_CMaPh}. Let us prove points \textbf{a)}
and \textbf{b)}. First, the critical points of $\mathcal{F}_{s}$
and $\mathcal{D}_{G}$ verify $s'(\omega_{c})=\beta\psi_{c}$ and
$\omega_{c}=G'\left(\beta\psi_{c}\right)$. Now using that $G$ is
the Legendre-Fenchel transform of $s$, if $s$ is differentiable,
we have $\left(s'\right)^{-1}=G'$. Thus the critical points of both
functionals are the same. 

Let us prove point \textbf{b)}\begin{eqnarray*}
\mathcal{F}_{s}[\omega] & = & -\int_{\mathcal{D}}d^{2}x\,\left[-s\left(\omega\right)+\beta\omega\psi\right]+\int_{\mathcal{D}}d^{2}x\,\frac{\beta}{2}\omega\psi\\
 & \geq & \int_{\mathcal{D}}d^{2}x\,\left[-G(\beta\psi)+\frac{\beta}{2}\omega\psi\right]=\mathcal{D}_{G}\left[\psi\right]\end{eqnarray*}
where we have used the definition of $G$, as the Legendre-Fenchel
transform of $s$, in order to prove the inequality. We now conclude
the proof of point \textbf{a)}. A direct computation gives $G(x)=x\left(s'\right)^{-1}(x)-s\left[\left(s'\right)^{-1}(x)\right]$.
Thus $G(\beta\psi_{c})=\beta\psi_{c}\omega_{c}-s\left(\omega_{c}\right)$.
This proves that an equality actually occurs in the preceding equations,
for the critical points : $\mathcal{F}_{s}[\omega_{c}]=\mathcal{D}_{G}\left[\psi_{c}\right]$.

We have thus established the relations between RSM equilibria and
the simpler Casimirs (CVP) and stream function (SFVP) variational
problems.

\paragraph*{Acknowledgment :}

I warmly thank J. Barré, T. Dauxois, F. Rousset and A. Venaille for
helpful comments and discussions.

\bibliography{FBouchet,Long_Range,Meca_Stat_Euler,Ocean,Experimental_2D_Flows,Euler_Stability}

\begin{thebibliography}{20}
\expandafter\ifx\csname natexlab\endcsname\relax\def\natexlab#1{#1}\fi
\expandafter\ifx\csname bibnamefont\endcsname\relax
  \def\bibnamefont#1{#1}\fi
\expandafter\ifx\csname bibfnamefont\endcsname\relax
  \def\bibfnamefont#1{#1}\fi
\expandafter\ifx\csname citenamefont\endcsname\relax
  \def\citenamefont#1{#1}\fi
\expandafter\ifx\csname url\endcsname\relax
  \def\url#1{\texttt{#1}}\fi
\expandafter\ifx\csname urlprefix\endcsname\relax\def\urlprefix{URL }\fi
\providecommand{\bibinfo}[2]{#2}
\providecommand{\eprint}[2][]{\url{#2}}

\bibitem[{\citenamefont{{Robert}}(1991)}]{Robert:1991_JSP_Meca_Stat}
\bibinfo{author}{\bibfnamefont{R.}~\bibnamefont{{Robert}}},
  \bibinfo{journal}{J. Stat. Phys.}  (\bibinfo{year}{1991}).

\bibitem[{\citenamefont{{Robert} and
  {Sommeria}}(1991)}]{Robert_Sommeria_1991JFM}
\bibinfo{author}{\bibfnamefont{R.}~\bibnamefont{{Robert}}} \bibnamefont{and}
  \bibinfo{author}{\bibfnamefont{J.}~\bibnamefont{{Sommeria}}},
  \bibinfo{journal}{J. Fluid Mech.} \textbf{\bibinfo{volume}{229}},
  \bibinfo{pages}{291} (\bibinfo{year}{1991}).

\bibitem[{\citenamefont{Miller}(1990)}]{Miller:1990_PRL_Meca_Stat}
\bibinfo{author}{\bibfnamefont{J.}~\bibnamefont{Miller}},
  \bibinfo{journal}{Phys. Rev. Lett.} \textbf{\bibinfo{volume}{65}},
  \bibinfo{pages}{2137} (\bibinfo{year}{1990}).

\bibitem[{\citenamefont{{Eyink} and
  {Sreenivasan}}(2006)}]{Eyink_Sreenivasan_2006_Rev_Modern_Physics}
\bibinfo{author}{\bibfnamefont{G.~L.} \bibnamefont{{Eyink}}} \bibnamefont{and}
  \bibinfo{author}{\bibfnamefont{K.~R.} \bibnamefont{{Sreenivasan}}},
  \bibinfo{journal}{Rev. Mod. Phys.} \textbf{\bibinfo{volume}{78}},
  \bibinfo{pages}{87} (\bibinfo{year}{2006}).

\bibitem[{\citenamefont{{Ellis} et~al.}(2002)\citenamefont{{Ellis}, {Haven},
  and {Turkington}}}]{EllisHavenTurkington:2002_Nonlinearity_Stability}
\bibinfo{author}{\bibfnamefont{R.~S.} \bibnamefont{{Ellis}}},
  \bibinfo{author}{\bibfnamefont{K.}~\bibnamefont{{Haven}}}, \bibnamefont{and}
  \bibinfo{author}{\bibfnamefont{B.}~\bibnamefont{{Turkington}}},
  \bibinfo{journal}{Nonlinearity} \textbf{\bibinfo{volume}{15}},
  \bibinfo{pages}{239} (\bibinfo{year}{2002}).

\bibitem[{\citenamefont{{Chavanis}}(2003)}]{Chavanis_Generalized_Entropy_2003}
\bibinfo{author}{\bibfnamefont{P.-H.} \bibnamefont{{Chavanis}}},
  \bibinfo{journal}{\pre} \textbf{\bibinfo{volume}{68}},
  \bibinfo{pages}{036108} (\bibinfo{year}{2003}).

\bibitem[{\citenamefont{{Chavanis}}(2005)}]{Chavanis_2005PhyD_Generalizedentro%
py}
\bibinfo{author}{\bibfnamefont{P.-H.} \bibnamefont{{Chavanis}}},
  \bibinfo{journal}{Physica D} \textbf{\bibinfo{volume}{200}},
  \bibinfo{pages}{257} (\bibinfo{year}{2005}).

\bibitem[{\citenamefont{{Arnold}}(1966)}]{Arnold_1966}
\bibinfo{author}{\bibfnamefont{V.~I.} \bibnamefont{{Arnold}}},
  \bibinfo{journal}{Izv. Vyssh. Uchebbn. Zaved. Matematika; Engl. transl.: Am.
  Math. Soc. Trans.} \textbf{\bibinfo{volume}{79}} (\bibinfo{year}{1966}).

\bibitem[{\citenamefont{{Holm} et~al.}(1985)\citenamefont{{Holm}, {Marsden},
  {Ratiu}, and {Weinstein}}}]{Holm_Marsden_Ratiu_Weinstein_1985PhysRev}
\bibinfo{author}{\bibfnamefont{D.~D.} \bibnamefont{{Holm}}},
  \bibinfo{author}{\bibfnamefont{J.~E.} \bibnamefont{{Marsden}}},
  \bibinfo{author}{\bibfnamefont{T.}~\bibnamefont{{Ratiu}}}, \bibnamefont{and}
  \bibinfo{author}{\bibfnamefont{A.}~\bibnamefont{{Weinstein}}},
  \bibinfo{journal}{Phys. Rep.} \textbf{\bibinfo{volume}{123}},
  \bibinfo{pages}{1} (\bibinfo{year}{1985}).

\bibitem[{\citenamefont{{Michel} and {Robert}}(1994)}]{Michel_Robert_1994JFM}
\bibinfo{author}{\bibfnamefont{J.}~\bibnamefont{{Michel}}} \bibnamefont{and}
  \bibinfo{author}{\bibfnamefont{R.}~\bibnamefont{{Robert}}},
  \bibinfo{journal}{J. Stat. Phys.} \textbf{\bibinfo{volume}{77}},
  \bibinfo{pages}{645} (\bibinfo{year}{1994}).

\bibitem[{\citenamefont{Bouchet and
  Sommeria}(2002)}]{Bouchet_Sommeria_2002_JFM_GRS}
\bibinfo{author}{\bibfnamefont{F.}~\bibnamefont{Bouchet}} \bibnamefont{and}
  \bibinfo{author}{\bibfnamefont{J.}~\bibnamefont{Sommeria}},
  \bibinfo{journal}{J. Fluid Mech.} \textbf{\bibinfo{volume}{464}},
  \bibinfo{pages}{465} (\bibinfo{year}{2002}).

\bibitem[{\citenamefont{Bouchet}(2001)}]{Bouchet_These}
\bibinfo{author}{\bibfnamefont{F.}~\bibnamefont{Bouchet}},
  \emph{\bibinfo{title}{M\'ecanique statistique des \'ecoulements
  g\'eophysiques}} (\bibinfo{publisher}{PHD, {\small Univ. J. Fourier}
  Grenoble}, \bibinfo{year}{2001}).

\bibitem[{\citenamefont{Bouchet and Dumont}(2003)}]{Bouchet_Dumont_Jupiter}
\bibinfo{author}{\bibfnamefont{F.}~\bibnamefont{Bouchet}} \bibnamefont{and}
  \bibinfo{author}{\bibfnamefont{T.}~\bibnamefont{Dumont}},
  \bibinfo{journal}{cond-mat/0305206}  (\bibinfo{year}{2003}).

\bibitem[{\citenamefont{{Ellis} et~al.}(2000)\citenamefont{{Ellis}, {Haven},
  and {Turkington}}}]{EllisHavenTurkington:2000_Inequivalence}
\bibinfo{author}{\bibfnamefont{R.~S.} \bibnamefont{{Ellis}}},
  \bibinfo{author}{\bibfnamefont{K.}~\bibnamefont{{Haven}}}, \bibnamefont{and}
  \bibinfo{author}{\bibfnamefont{B.}~\bibnamefont{{Turkington}}},
  \bibinfo{journal}{J. Stat. Phys.} \textbf{\bibinfo{volume}{101}}
  (\bibinfo{year}{2000}).

\bibitem[{\citenamefont{{Dauxois} et~al.}(2002)\citenamefont{{Dauxois},
  {Ruffo}, {Arimondo}, and {Wilkens}}}]{DRAW:2002_Houches}
\bibinfo{editor}{\bibfnamefont{T.}~\bibnamefont{{Dauxois}}},
  \bibinfo{editor}{\bibfnamefont{S.}~\bibnamefont{{Ruffo}}},
  \bibinfo{editor}{\bibfnamefont{E.}~\bibnamefont{{Arimondo}}},
  \bibnamefont{and}
  \bibinfo{editor}{\bibfnamefont{M.}~\bibnamefont{{Wilkens}}}, eds.,
  \emph{\bibinfo{title}{{Dynamics and Thermodynamics of Systems With Long Range
  Interactions}}} (\bibinfo{year}{2002}).

\bibitem[{\citenamefont{Bouchet and
  Barr\'e}(2005)}]{Bouchet_Barre_2005_JSP_Classification}
\bibinfo{author}{\bibfnamefont{F.}~\bibnamefont{Bouchet}} \bibnamefont{and}
  \bibinfo{author}{\bibfnamefont{J.}~\bibnamefont{Barr\'e}},
  \bibinfo{journal}{J. Stat. Phys.} \textbf{\bibinfo{volume}{118 5/6}},
  \bibinfo{pages}{1073} (\bibinfo{year}{2005}).

\bibitem[{\citenamefont{{Smith} and
  {O'Neil}}(1990)}]{Smith_ONeil_Physics_Fluids_1990_Inequivalence}
\bibinfo{author}{\bibfnamefont{R.~A.} \bibnamefont{{Smith}}} \bibnamefont{and}
  \bibinfo{author}{\bibfnamefont{T.~M.} \bibnamefont{{O'Neil}}},
  \bibinfo{journal}{Physics of Fluids B} \textbf{\bibinfo{volume}{2}},
  \bibinfo{pages}{2961} (\bibinfo{year}{1990}).

\bibitem[{\citenamefont{{Kiessling}}(1993)}]{Kiessling_1993_PointVortex_CommPu%
reApplMaths}
\bibinfo{author}{\bibfnamefont{M.~K.~H.} \bibnamefont{{Kiessling}}},
  \bibinfo{journal}{Comm. Pure Appl. Math.} \textbf{\bibinfo{volume}{47}},
  \bibinfo{pages}{27} (\bibinfo{year}{1993}).

\bibitem[{\citenamefont{{Caglioti} et~al.}(1995)\citenamefont{{Caglioti},
  {Lions}, {Marchioro}, and
  {Pulvirenti}}}]{CagliotiLMP:1995_CMP_II(Inequivalence)}
\bibinfo{author}{\bibfnamefont{E.}~\bibnamefont{{Caglioti}}},
  \bibinfo{author}{\bibfnamefont{P.~L.} \bibnamefont{{Lions}}},
  \bibinfo{author}{\bibfnamefont{C.}~\bibnamefont{{Marchioro}}},
  \bibnamefont{and}
  \bibinfo{author}{\bibfnamefont{M.}~\bibnamefont{{Pulvirenti}}},
  \bibinfo{journal}{Commun. Math. Phys.} \textbf{\bibinfo{volume}{174}},
  \bibinfo{pages}{229} (\bibinfo{year}{1995}).

\bibitem[{\citenamefont{{Wolansky} and
  {Ghil}}(1998)}]{Wolansky_Ghil_1998_CMaPh}
\bibinfo{author}{\bibfnamefont{G.}~\bibnamefont{{Wolansky}}} \bibnamefont{and}
  \bibinfo{author}{\bibfnamefont{M.}~\bibnamefont{{Ghil}}},
  \bibinfo{journal}{Commun. Math. Phys.} \textbf{\bibinfo{volume}{193}},
  \bibinfo{pages}{713} (\bibinfo{year}{1998}).

\end{thebibliography}

\end{document}